%% file: main.tex
\documentclass[10pt,conference]{IEEEtran}
\IEEEoverridecommandlockouts

\input{setup}

\usepackage{cite}
\usepackage{amsmath,amssymb,amsfonts}
\usepackage{graphicx}
\usepackage{textcomp}
\usepackage{xcolor}
\usepackage{colortbl}
\usepackage{todonotes}
\def\BibTeX{{\rm B\kern-.05em{\sc i\kern-.025em b}\kern-.08em
    T\kern-.1667em\lower.7ex\hbox{E}\kern-.125emX}}
\begin{document}

\title{Early Detection of Performance Regressions by Bridging Local Performance Data and Architectural Models}

\author{
\IEEEauthorblockN{
Lizhi Liao\IEEEauthorrefmark{1}, 
Simon Eismann\IEEEauthorrefmark{2}, 
Heng Li\IEEEauthorrefmark{3},
Cor-Paul Bezemer\IEEEauthorrefmark{4},
Diego Elias Costa\IEEEauthorrefmark{5}, 
André van Hoorn\IEEEauthorrefmark{6}, 
Weiyi Shang\IEEEauthorrefmark{1}
}
\IEEEauthorblockA{
\IEEEauthorrefmark{1}University of Waterloo,
\IEEEauthorrefmark{2}University of Würzburg,
\IEEEauthorrefmark{3}Polytechnique Montréal, 
\IEEEauthorrefmark{4}University of Alberta, \\
\IEEEauthorrefmark{5}Concordia University, 
\IEEEauthorrefmark{6}University of Hamburg \\
\IEEEauthorrefmark{1}\{lizhi.liao, wshang\}@uwaterloo.ca,
\IEEEauthorrefmark{2}simon.eismann@uni-wuerzburg.de,
\IEEEauthorrefmark{3}heng.li@polymtl.ca, \\
\IEEEauthorrefmark{4}bezemer@ualberta.ca,
\IEEEauthorrefmark{5}diego.costa@concordia.ca, 
\IEEEauthorrefmark{6}andre.van.hoorn@uni-hamburg.de
}
}

\maketitle

\begin{abstract}
\input{tex/Abstract}
\end{abstract}


\input{tex/Introduction}
\input{tex/Background}
\input{tex/Approach}
\input{tex/Evaluation}
\input{tex/Results}
\input{tex/RelatedWork}
\input{tex/Threats}
\input{tex/Conclusion}

\section*{Acknowledgment}
\revised{}{This research was conducted by the SPEC RG DevOps Performance Working Group.\footnote{\url{https://research.spec.org/devopswg}}}

\bibliographystyle{IEEETrans}
\bibliography{bibliography}

\end{document}

%% file: setup.tex
\usepackage{booktabs}
\usepackage{makecell}
\usepackage{algorithmicx}
\usepackage{algpseudocode}
\usepackage{amsmath}
\usepackage[linesnumbered,ruled,vlined]{algorithm2e}
\usepackage{algpseudocode}
\usepackage{tcolorbox}
\usepackage{multirow}
\usepackage{threeparttable}
\usepackage{array}
\usepackage{float}
\usepackage{subcaption}
\usepackage{graphicx}
\usepackage{url}

\newcommand{\revised}[2]{#2}

\usepackage{fancybox}

\newcommand{\findingboxx}[1]{
\begin{center}
\begin{tcolorbox}[colback=gray!11,%
                  colframe=black,%
                  boxrule=0.2mm,
                  width=0.48\textwidth,%
                  arc=.8mm, auto outer arc,
                 ]
 #1
\end{tcolorbox}
\end{center}}

%% file: tex/Abstract.tex
During software development, developers often make numerous modifications to the software to address existing issues or implement new features. However, certain changes may inadvertently have a detrimental impact on the overall system performance.
To ensure that the performance of new software releases does not degrade (i.e., absence of performance regressions), existing practices rely on system-level performance testing, such as load testing, or component-level performance testing, such as microbenchmarking, to detect performance regressions.
However, performance testing for the entire system is often expensive and time-consuming, posing challenges to adapting to the rapid release cycles common in modern DevOps practices. In addition, system-level performance testing cannot be conducted until the system is fully built and deployed.
On the other hand, component-level testing focuses on isolated components, neglecting overall system performance and the impact of system workloads.
In this paper, we propose a novel approach to early detection of performance regressions by bridging the local performance data generated by component-level testing and the system-level architectural models.
Our approach uses local performance data to identify deviations at the component level, and then propagate these deviations to the architectural model. We then use the architectural model to predict regressions in the performance of the overall system.
In an evaluation of our approach on two representative open-source benchmark systems, we show that it can effectively detect end-to-end system performance regressions from local performance deviations with different intensities and under various system workloads.
More importantly, our approach can detect regressions as early as in the development phase, in contrast to existing approaches that require the system to be fully built and deployed. Our approach is lightweight and can complement traditional system performance testing when testing resources are scarce.

%% file: tex/Introduction.tex
\section{Introduction}
\label{sec:introduction}

Performance is a critical aspect of the quality of service (QoS) of software systems. It is important to ensure that software consistently delivers optimal performance after each new release (i.e., absence of performance regressions)~\cite{DBLP:books/daglib/0023085, DBLP:journals/tse/WeyukerV00, DBLP:conf/issta/Jiang10, bondi2014foundations, smith2002performance}.
However, throughout the development cycle, developers may implement various modifications to the software to implement new features or address existing issues, some of which could potentially impact the performance of the system adversely~\cite{8946099, DBLP:journals/tse/ChenSS22, DBLP:conf/icse/ChenL00LL22}.
Such performance regressions can result in higher resource consumption (e.g., excessive memory or CPU usage), increased response time, or even field failures, thereby causing significant financial and reputation losses~\cite{DBLP:conf/icse/WoodsideFP07, website_load_time}.
For instance, according to a recent report~\cite{website_load_time}, even a mere two-second difference in the website response time can drastically decrease user satisfaction, causing the bounce rate to surge from 9\% to 38\%. Ultimately, this could result in a remarkable loss of market share and revenue.
Therefore, it is crucial to detect and resolve any performance regressions before deploying the system.

In practice, traditional system-level performance testing is widely adopted to prevent potential performance regressions from sneaking into production~\cite{DBLP:conf/icst/GaoJBL16, DBLP:conf/apsec/NguyenAJHNF11, DBLP:conf/icse/MalikHH13}.
Existing practices involve running field-like workloads for an extended period of time (from hours to days) to exercise both the old and the new versions of the system in an in-house performance testing environment.
During testing, a large number of performance metrics and execution logs are generated, and performance analysts collect and analyze this information from both old and new versions to determine the existence of performance regressions~\cite{DBLP:conf/icst/GaoJBL16}.
However, performance testing can be expensive, especially for large-scale systems, requiring significant resources and time to set up the environment and execute the performance tests~\cite{DBLP:conf/kbse/ChenSHWL19}.
Furthermore, such a process is often conducted late in the software development and release cycle, i.e., after the system is built or even integrated, making it challenging and laborious to diagnose and address the performance regressions at such a late stage.

To tackle this challenge, there has been a growing interest in leveraging component-level (e.g., function or class) performance information to detect performance regressions in software systems.
These approaches~\cite{DBLP:conf/icsm/ChenS17, DBLP:conf/wosp/HorkyLMST15, DBLP:conf/icse/DingCS20, DBLP:conf/ssiri/KimCW09, DBLP:conf/kbse/ReicheltKH19, DBLP:conf/msr/LaaberL18, jangali2022automated} propose to leverage unit tests (e.g., JUnit tests) or microbenchmarks (e.g., Java Microbenchmark Harness) to collect the corresponding performance metrics, such as execution time or throughput.
Through conducting statistical analysis on the performance metrics collected from the test runs before and after code changes, developers can identify any remarkable performance improvement or regressions in individual components of the system.
Such approaches are relatively lightweight to execute and avoid the expensive resources and time required for running traditional performance tests on the entire system.
However, due to the unique nature of component-level tests, they typically do not explain the system-level performance well~\cite{DBLP:conf/kbse/ChenSHWL19}.
In particular, the performance changes of components may be propagated to the system very differently. For instance, a function executed once may have a negligible performance impact whereas a function executed in a long loop may have a much more significant impact. In addition, it is also challenging to consider realistic workloads in the component-level testing.

In this paper, we propose a novel approach to early detection of performance regressions by bridging the local performance data generated by component-level testing and the architectural models of the system.
After developers make code changes during development, we first collect local performance information by running component-level tests and then identify performance deviations at the component level. Afterward, we propagate these component-level performance deviations to the architectural model. Finally, we leverage the architectural model to predict performance regressions of the entire system.
To evaluate the effectiveness of our approach, we conduct experiments on two representative open-source benchmark systems, i.e., TeaStore~\cite{KiEiScBaGrKo2018-MASCOTS-TeaStore} and TrainTicket~\cite{DBLP:journals/tse/ZhouPXSJLD21}.
In the experimental results, we find that by bridging local performance data and architectural models, our approach can effectively detect end-to-end system performance regressions with different intensities of local performance deviations and can maintain its effectiveness when the system experiences various workloads. 
\revised{RBC13}{The experimental results also demonstrate that our approach can assist developers in identifying and addressing performance regressions as early as the software development phase and can be adapted into the rapid software development and release practices (e.g., DevOps) to complement traditional system performance testing in resource-constrained scenarios.}

This paper makes the following main contributions:
\begin{itemize}
    \item We develop a novel approach to the early detection of performance regressions by bridging the local performance testing data and the system architectural models.

    \item Our approach is lightweight and can be integrated into fast-paced software development and release practices (e.g., DevOps) to complement traditional system performance testing in situations where testing resources are scarce.
\end{itemize}

\noindent\textbf{Paper organization.} 
Section~\ref{sec:background} discusses the background of our work.
Section~\ref{sec:approach} outlines our approach to detecting system performance regressions.
Section~\ref{sec:evaluation} introduces the evaluation setup.
Section~\ref{sec:results} presents the evaluation analysis and results.
Section~\ref{sec:relatedwork} discusses the prior related research.
The threats to the validity of our work are discussed in Section~\ref{sec:threats}.
Finally, Section~\ref{sec:conclusion} concludes our work.

%% file: tex/Background.tex
\section{Background}
\label{sec:background}

In this section, we introduce the background of our study, including software performance testing and model-based performance analysis.

\subsection{Software performance testing}
Software performance testing evaluates a software system's performance under various conditions to ensure the system meets the specified performance requirements and operates efficiently. Software performance testing can be conducted at both the system and component levels.

At the \underline{system level}, performance testing evaluates the overall performance of the entire software system under a workload similar to real-world scenarios.
It simulates user interactions and workload scenarios to assess performance metrics such as response times, throughput, and resource utilization of the system.
Typically, there are four phases in system performance testing: 1) defining a system workload (e.g., Apache JMeter HTTP(S) test script ~\cite{halili2008apache}), 2) preparing an in-house testing environment, 3) executing the performance tests, and 4) analyzing the testing results.
System performance testing has demonstrated its effectiveness in assisting developers in determining compliance with performance goals, identifying system bottlenecks, and detecting performance regressions~\cite{DBLP:conf/icst/GaoJBL16}. However, since it involves the entire software system, and modern software systems are often large in scale and highly complex, the testing process can be extremely time- and resource-consuming.

On the other hand, at the \underline{component level}, performance testing focuses on evaluating the performance of individual software components or modules in isolation. These components can be different functions or classes that can be independently tested.
Component-level performance testing involves developers writing tests (microbenchmarks) targeting specific components of concern with performance unit testing frameworks, such as JMH, JunitPerf, and ContiPerf, or directly using JUnit to test the performance combined with functional tests~\cite{DBLP:journals/tse/Costa:JMH,DBLP:journals/ese/LaaberSL19}.
This type of performance testing is lightweight to execute and effective to uncover performance issues at the component level and ensure components meet performance requirements individually~\cite{DBLP:conf/icse/DingCS20}.
However, it often does not explain the system-level performance well~\cite{DBLP:conf/kbse/ChenSHWL19}, since performance changes of a component may be propagated to the system very differently (e.g., the function executed once versus executed within a long loop).
Furthermore, it is also challenging to consider realistic workloads in the component-level testing.

\subsection{Model-based performance analysis}
\begin{figure}[!t]
\centering
\begin{subfigure}{0.47\textwidth}
\centering
\includegraphics[width=0.93\textwidth]{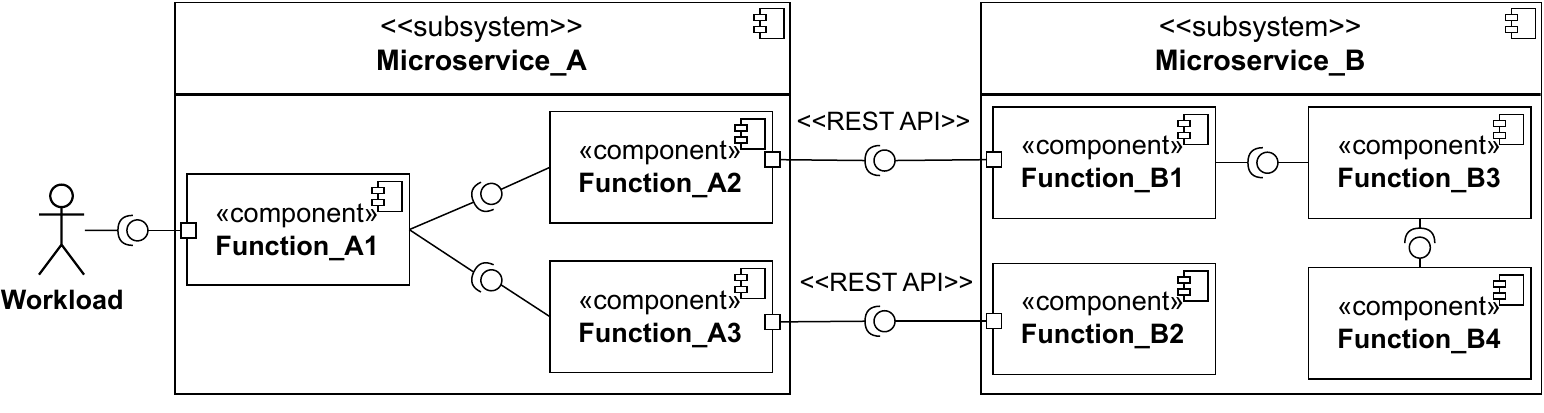}
\caption{An example software system with two subsystems}
\label{fig:example_system}
\end{subfigure}
\\[1ex]
\begin{subfigure}{0.47\textwidth}
\centering
\includegraphics[width=0.83\textwidth]{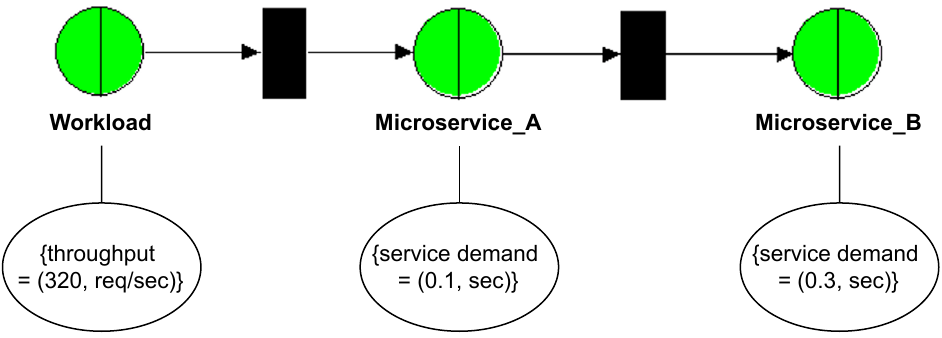}
\caption{The corresponding architectural (QPN) model (the green dots indicate service queueing places, the underneath circles indicate corresponding performance parameters, and the black boxes indicate transitions between queueing places)}
\label{fig:example_model}
\end{subfigure}
\vspace{-.3em}
\caption{An example software system with two subsystems and its corresponding architectural model}
\vspace{-1.7em}
\label{fig:example_system_model}
\end{figure}

Model-based performance analysis is a methodology that utilizes analytical models to predict and assess the performance of software systems~\cite{DBLP:journals/tse/BalsamoMIS04}.
\revised{RBC1, RBC7}{Model-based performance analysis involves constructing abstract models of the system, describing its components, interactions, and resource utilization, and then employing mathematical and computational techniques for analysis and simulation~\cite{DBLP:journals/jss/BeckerKR09, DBLP:conf/wosp/BeckerKR07}.}
Models can take various forms, such as queueing network (QN)~\cite{kleinrock1975queuing,kant1992introduction, DBLP:journals/tc/WoodsideNPM95,rolia1995method}, layered queueing network (LQN)~\cite{DBLP:journals/tse/FranksOWDD09, DBLP:journals/entcs/ShoaibD11, DBLP:conf/cmg/TiwariM06}, or queueing Petri nets (QPN)~\cite{baccelli1994annotated, DBLP:conf/pnpm/Bause93, bause1993qn, bause1997integrating}, to capture different aspects and characteristics of the system.
\revised{RAC5}{Figure~\ref{fig:example_system} shows an illustrative example of a software system with two subsystems and Figure~\ref{fig:example_model} presents the corresponding architectural model of the system.}
Model-based performance analysis provides an efficient and effective means for developers and system designers to better understand and optimize the performance characteristics of software systems.
Through modeling, developers can identify potential performance bottlenecks, predict system performance under various workload conditions, and make design optimizations to enhance system performance~\cite{DBLP:journals/tse/Kounev06}.
However, when dealing with modern systems that are typically large-scale and complex, accurately modeling the performance of the entire system is challenging, leading to the adoption of high-level modeling representations, such as modeling at the subsystem or service levels~\cite{DBLP:conf/cmg/KounevB03} as illustrated in Figure~\ref{fig:example_system_model}. While these high-level models offer a macroscopic understanding of the overall system performance, they also constrain the model's ability to capture nuanced performance changes at the system's finer levels (e.g., functions).

In this paper, we propose a novel cost-effective approach to detecting system performance regressions during the development phase by bridging local performance data generated from component-level performance testing and architectural models.
Our approach allows developers to understand earlier that there is a performance regression, instead of running expensive system performance testing after the system is built or even integrated.
We present our approach in detail in the next section.

%% file: tex/Approach.tex
\section{Approach}
\label{sec:approach}

\begin{figure*}[!t]
	\centering
	\includegraphics[width=0.97\textwidth]{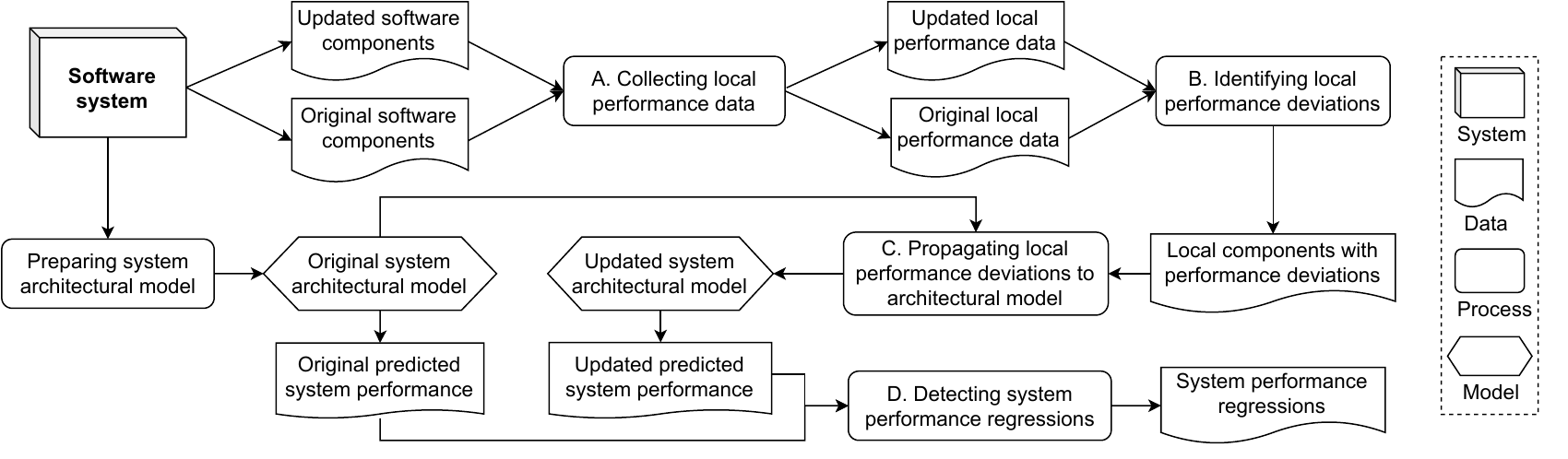}
 \vspace{-.5em}
	\centering \caption{An overview of our approach for system performance regression detection by bridging local performance data and architectural models}
    \vspace{-1.7em}
	\label{fig:approach_overview}
\end{figure*}

In this section, we present our approach for detecting system performance regressions by bridging local performance data and architectural models. 
Figure~\ref{fig:approach_overview} gives an overview of our approach, and each subsection corresponds to a step in the figure. 
To ease the illustration of our approach, we show a running example of a software system with two subsystems, each with three and four functions, as shown in Figure~\ref{fig:example_system}. However, the number of subsystems and functions is much larger in our evaluation and real-life scenarios. 
\revised{RCC1}{Our approach is designed to be lightweight and can be seamlessly integrated into the software development process, such as the CI/CD pipeline.
In particular, after developers make changes to the software (e.g., after a commit), our approach aims to provide developers with timely feedback on whether these changes will cause system performance regressions or not, without the need for running expensive system-level performance testing.
}

\noindent\textbf{Preparing system architectural model.}
Our approach relies on the architectural model to detect performance regressions. 
The model is typically available as part of the software artifact~\cite{DBLP:journals/jss/Heinrich20} and can be directly utilized for our approach. 
\revised{RAC1}{As an alternative when there is no available model, we can also recover the architectural model through relatively low-frequency system-level performance testing, which has been extensively discussed in prior studies~\cite{DBLP:journals/entcs/ShoaibD11, DBLP:journals/tse/Kounev06, DBLP:journals/jss/CortellessaPET22}.}
In particular, compared to component-level performance testing that typically occurs after each commit (alongside unit tests in the CI pipeline), system-level performance testing is often conducted only once at the beginning of a major release~\cite{DBLP:conf/icst/GaoJBL16}.
It is also worth noting that our approach does not necessarily depend on system performance testing and specific types of architectural models. Instead, our approach primarily focuses on bridging the local performance data and architectural models to detect performance regressions, assuming the availability of the system architectural model.

\subsection{Collecting local performance data}
\label{approach_collect_local}
In the first step, we collect the performance data of local components (i.e., functions) before and after developers make changes to the software during development.
In particular, we run the component-level tests of the software, e.g., unit tests or JMH microbenchmarks, for the performance testing of the local components. We execute these test cases on both the original and the updated version of the code. 
\revised{RAC2, RBC2}{Each test is executed with a total of 30 iterations; in each iteration, we compile, deploy, and test the system from scratch, to minimize the impact of fluctuations and outliers in the performance data (potentially caused by environmental noise, cache effects, system randomness, or flaky tests). This also follows the performance testing practices that have been done in prior work~\cite{delgado2021performance}.}
Furthermore, during testing, we actively utilize application performance monitoring tools to monitor and collect the software system’s runtime behavior (i.e., the interactions of local components – functions), and the corresponding performance data (i.e., the execution time of each component).

\subsection{Identifying local performance deviations}
\label{approach_identify_local}
We analyze the local performance data collected in the previous step to estimate the performance deviation (in execution time) of local software components.
We first conduct Wilcoxon rank-sum statistical tests~\cite{wilcoxon1992individual} to compare the performance data distributions before and after the update. We run the statistical test at the 5\% level of significance, and the null hypothesis is that there exists no statistically significant difference between the performance of the original and updated versions of the component. If the p-value of the test is not greater than 0.05, we would reject the null hypothesis and favor the alternative hypothesis, i.e., there is a statistically significant difference.
However, recognizing that statistical tests solely reveal the existence of differences without quantifying their magnitude, we complement this statistical analysis by calculating Cliff's Delta effect size~\cite{cliff2014ordinal} to determine the magnitude of the observed differences. 
We employ the widely adopted thresholds for Cliff's Delta effect size, as provided in prior research ~\cite{romano2006appropriate}.
It is worth noting that the choices of the Wilcoxon rank-sum test and Cliff’s Delta effect size are deliberate since they do not assume a specific distribution for the performance data.

In the cases where statistical analysis indicates significant differences with an effect size larger than negligible, we further propagate the local performance deviation to the system architectural model. To do so, we calculate the mean difference (MD) of local performance before and after updates. A positive value indicates decreased performance (i.e., longer execution time) in the updated version, while a negative value suggests performance improvement (i.e., shorter execution time).

\subsection{Propagating local performance deviations to architectural model}

\begin{figure}[!t]
\centering
\begin{subfigure}{0.47\textwidth}
\centering
\includegraphics[width=0.73\textwidth]{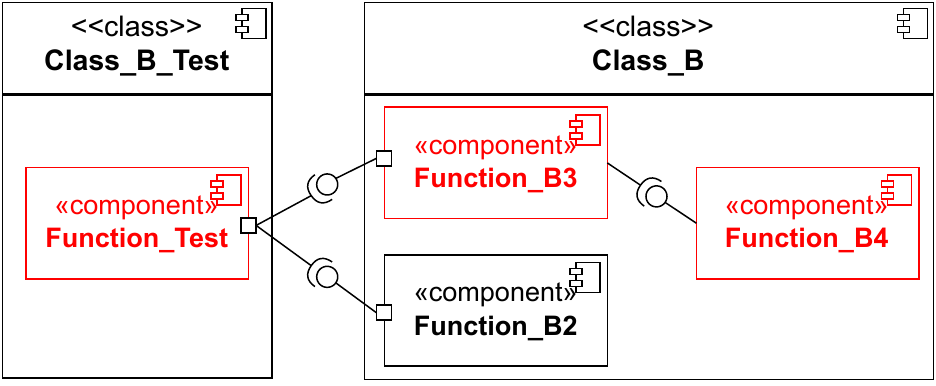}
\vspace{-.5em}
\caption{Extracting local performance deviations}
\label{fig:local_perf_regression}
\end{subfigure}
\\[1ex]
\begin{subfigure}{0.47\textwidth}
\includegraphics[width=0.97\textwidth]{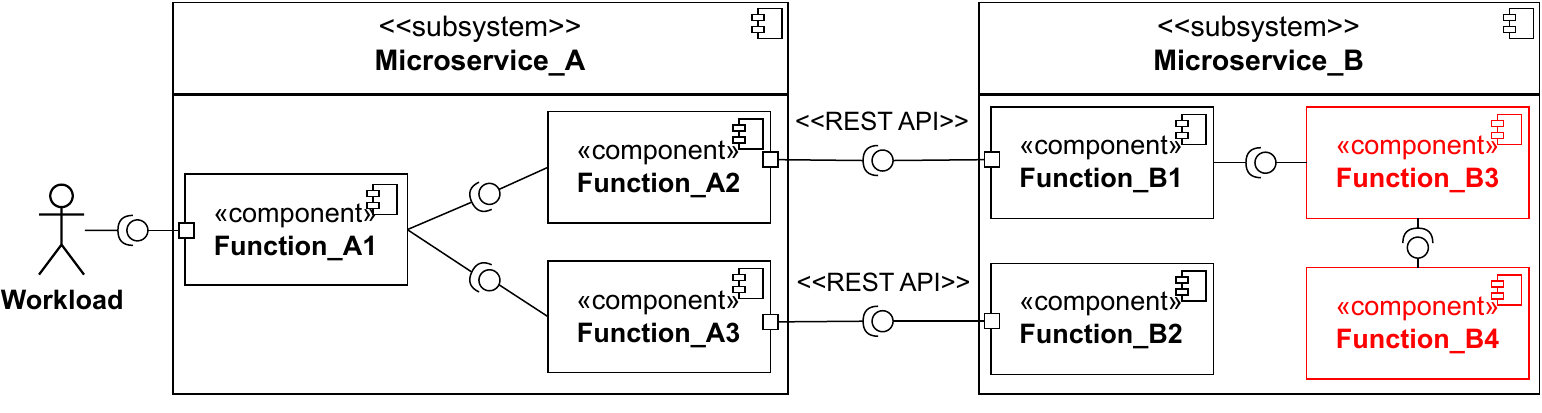}
\vspace{-.5em}
\caption{Mapping local performance deviations to the system level via maximum common subgraph analysis}
\label{fig:max_sub_graph}
\end{subfigure}
\\[1ex]
\begin{subfigure}{0.47\textwidth}
\includegraphics[width=0.97\textwidth]{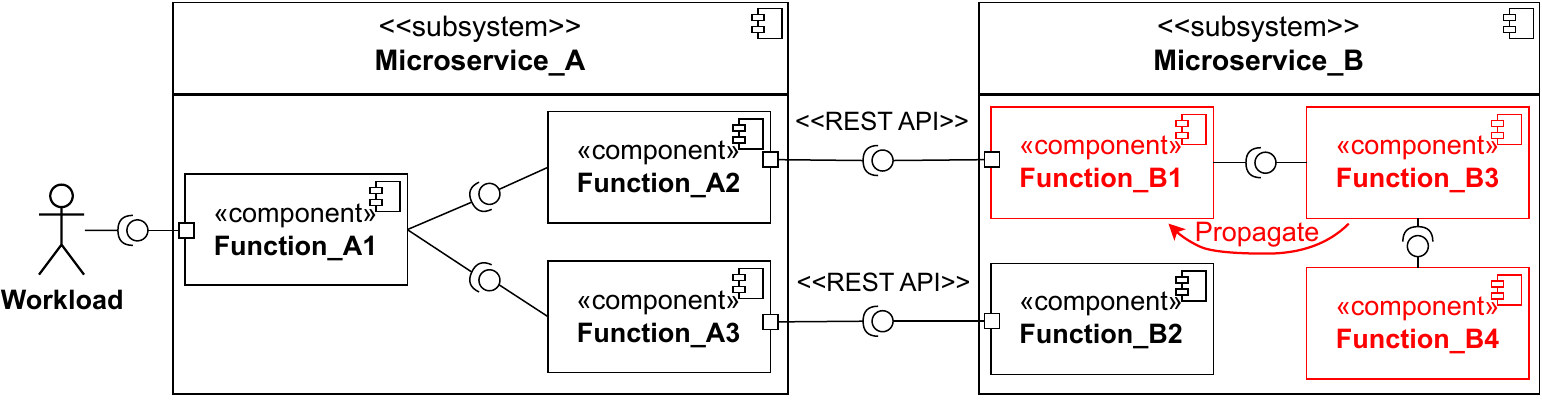}
\vspace{-.5em}
\caption{Propagating performance deviations to the top level of the subsystem}
\label{fig:propagating}
\end{subfigure}
\\[1ex]
\begin{subfigure}{0.47\textwidth}
\centering
\includegraphics[width=0.83\textwidth]{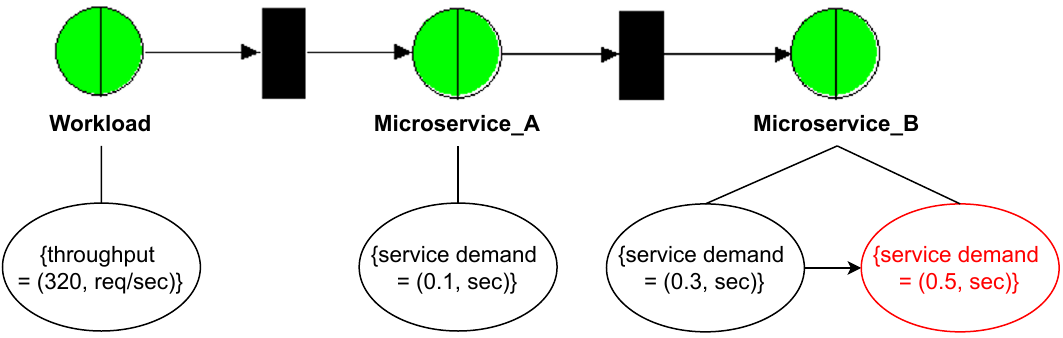}
\vspace{-.5em}
\caption{Updating architectural model}
\label{fig:updating_model}
\end{subfigure}
\vspace{-.3em}
\caption{An illustrative example of propagating local performance deviations to architectural model (for subfigures (a), (b), and (c), the red boxes indicate components with performance deviations, while for subfigure (d), the red box indicates updated performance parameter in the architectural model)}
\vspace{-1.7em}
\end{figure}

With the identified local software components that suffer from deviated performance, intuitively, one may utilize the local performance deviations and the number of times the components are called as a proxy to determine system performance regressions.
However, such a naïve approach may overlook the crucial impact of the system's complexity and dynamics, such as resource contention, queue wait times, and various workloads.
Therefore, the architectural model is used to detect performance regressions, especially when the system is under various workloads.
In this step, we first analyze the local performance data collected from the prior step (cf. Section~\ref{approach_collect_local}) to establish a dependency graph to represent the structure and performance of the local components. Such a graph is a directed acyclic graph, with nodes representing the local components, and edges representing their interaction dependencies (e.g., function calls). Performance information is represented as node attributes.
We then extract a subgraph with local components that have deviated performance (identified from the previous step in Section~\ref{approach_identify_local}) within the local dependency graph and mark them accordingly, as depicted in Figure~\ref{fig:local_perf_regression}.

Afterward, we map the local performance deviations to the corresponding system level through graph mapping.
In particular, we first identify common substructures shared between the extracted local graph and the existing system graph by leveraging the ISMAGS algorithm~\cite{houbraken2014index}, which finds the maximum common subgraph of two graphs.
We then align the performance deviation (i.e., MD) from the local level to the corresponding system level.
Figure~\ref{fig:max_sub_graph} shows an example of mapping the local performance deviation marked in Figure~\ref{fig:local_perf_regression} to the corresponding system level.

Since system architectural models typically operate at a relatively higher level, such as the subsystem level, merely understanding the performance deviations at the lower (component) level is insufficient. Therefore, we further employ a bottom-up strategy to propagate performance deviations upwards through the hierarchical dependency layers of the system.
We begin this process from the identified components of concern and extend it until it reaches the top-level components of that subsystem, since these components encompass the performance of their invoked components and constitute the performance of the subsystem.
As illustrated in the example in Figure~\ref{fig:propagating}, after we successfully map the local performance deviation to the system level for Function\_B3 and Function\_B4, we further propagate to the top-level component Function\_B1.

Finally, we calculate the overall performance of the subsystem by summing up the top-level component performance (e.g., Function\_B1 and Function\_B2 in Figure~\ref{fig:propagating}) and calculate the deviation (i.e., relative difference) between the overall performance of the subsystem before and after the changes.
With this deviation in the overall performance of the subsystem, we then update the corresponding performance parameters within the architectural model. 
\revised{RCC8}{For example, in Figure~\ref{fig:updating_model}, we propagate the local performance deviation to the architectural model by updating the service demand parameter for Microservice\_B from 0.3 seconds to 0.5 seconds.}

\subsection{Detecting system performance regressions}
The last step of our approach is to identify system performance regressions by comparing the original architectural model and the model updated in the last step.
We first use both models to get the original and updated predicted performance, including the resource utilization and response time.
We then utilize a statistical analysis approach similar to the one used for identifying local performance deviation in Section~\ref{approach_identify_local}. In particular, we utilize the Wilcoxon rank-sum test and Cliff’s Delta effect size, to determine whether there is a significant difference between the two performance distributions and the magnitude of the difference.
If the p-value is smaller than 0.05 and the effect size is larger than negligible, we would consider that the changes to the software may lead to system performance regressions.

%% file: tex/Evaluation.tex
\section{Evaluation Setup}
\label{sec:evaluation}

\revised{RAC3}{To evaluate the effectiveness of our approach, we conduct case studies on two prevalent open-source systems (i.e., TeaStore and TrainTicket) with injected performance regressions\footnote{Our evaluation setup, scripts, and results are shared online via \url{https://doi.org/10.5281/zenodo.13135986} as a replication package.}.
Our selected subject systems are well-established benchmark systems that provide detailed documentation, abundant system tests, and a certain extent of representativeness with real-world systems regarding scale and complexity. In addition, these
systems have been widely used and studied in prior software performance engineering studies~\cite{DBLP:journals/jss/CortellessaPET22, DBLP:journals/tse/LiaoCLZSSTS22, DBLP:conf/sigsoft/SokolowskiWS21}.}
In this section, we present the details about the subject systems, evaluation environment, performance regressions, component-level and system-level performance test design, architectural models, and evaluation scenario design.

\subsection{Subject systems}
\label{subject_systems}

\revised{RAC3}{
\textbf{TeaStore}~\cite{KiEiScBaGrKo2018-MASCOTS-TeaStore} is a microservice reference and benchmark application that serves as a simulated web store. 
As a distributed microservice application, TeaStore consists of five distinct services plus a separate service registry.
We choose TeaStore as a subject since it is positioned as a benchmark application and widely utilized for evaluating various performance aspects~\cite{KiEiScBaGrKo2018-MASCOTS-TeaStore}, including performance modeling, auto-scaling, and energy-efficiency analysis.
The system comprises approximately 33k lines of source code and 400 files.  
}

\revised{RAC3, RBC10}{
\textbf{TrainTicket}~\cite{DBLP:journals/tse/ZhouPXSJLD21} is a web train ticket booking system that provides various typical train ticket booking functionalities, such as ticket inquiry, reservation, payment, and rebooking.
This system is a benchmark system and also adopts a microservice architecture, containing up to 41 microservices. 
The system comprises approximately 289k lines of source code and 2k files. 
}

\subsection{Evaluation environment}
The evaluation of our subject systems is conducted on two separate machines interconnected within the same internal network. Each machine has a configuration of Intel Core i7-6700 CPU, 32GB of RAM, and operates on Ubuntu 22.04.3 LTS.
The first machine functions as the application server to host the subject system, while the second machine serves to generate workloads (via JMeter load driver~\cite{halili2008apache}) to simulate real-world users interacting with the system.
We leverage Podman container (v4.6.2) and Podman-Compose orchestration (v1.0.6) for the deployment and management of our subject systems, where each container runs a particular microservice. It is worth noting that we deliberately restrict the CPU usage for each service container to avoid resource contention among containers.
We utilize the logs generated by the load generator to collect the response time of each request and leverage \emph{podman stats} to gather the resource (CPU) usage statistics for each of the containers. 
Furthermore, we have instrumented the subject systems with Kieker~\cite{van2012kieker} to collect the execution time of each function and their interactions. Kieker offers comprehensive dynamic analysis capabilities at runtime with low overhead and does not require modifying the source code.

\subsection{Performance regressions}
\label{performance_regressions}
By checking the development history (i.e., commit logs, issue tracking, and comments) of our subject systems, we did not find any historical performance regressions of specific commits or versions. Therefore, we opt to manually inject performance regressions in the source code of the subject systems for the evaluation of our approach.
\revised{RBC8}{We consider injecting performance regression by adding busy waiting in the source code to slow down the software execution. Although real-world performance regressions can differ, busy waiting has been used in previous studies~\cite{DBLP:conf/msr/LaaberL18, jangali2022automated, delgado2021performance} to simulate additional heavy-weight operations.
}
To mitigate the bias introduced by injecting performance regressions at one specific location in the software, we systematically examine the source code of our subject systems and arbitrarily identify three distinct locations from different subsystems and different components for regression injection, denoted as L1, L2, and L3.

Furthermore, we also investigate the performance regressions with various intensities, i.e., different waiting time lengths.
According to prior work on evaluating software microbenchmark suites, Laaber and Leitner~\cite{DBLP:conf/msr/LaaberL18} observe that slowdowns exceeding 50\% (i.e., 1.5x slower) in most Java projects can often be reliably detected.
\revised{RAC4}{Inspired by this finding, at each of the selected locations in the source code, we respectively inject busy waiting with three distinct time lengths, including 10\%, 50\%, and 250\% of the component's original execution time, and denote them as Low, Medium, and High.
This process would yield a total of nine distinct versions per subject system, each of which incorporates the performance regression of a specific intensity and at a specific location. 
In addition, we measure the performance of each version in isolation to prevent interference between different injections.}

\subsection{Component-level and system-level performance test design}
\label{test_design}
Our evaluation relies on component-level performance tests to understand the impact of the injected performance regressions at the component level and further propagate such impact to the architectural model.
For both of our subject systems, although they lack existing performance microbenchmarks, there exists a large number of unit tests, i.e., TeaStore has 163 test cases and TrainTicket has 628 test cases. 
In addition, prior studies~\cite{jangali2022automated, DBLP:conf/icse/DingCS20, DBLP:journals/ese/LaaberSL19}, have pointed out that unit tests can be a good proxy for performance tests (if there are no real performance tests). 
Therefore, we leverage these existing unit tests to facilitate the performance testing of the local components of our subject systems.
We run each test a total of 30 times to minimize environmental noise and system randomness, similar to what was done in prior work~\cite{delgado2021performance}. 

While our approach does not inherently require system-level performance tests (cf. Section~\ref{sec:approach}), our study still relies on these tests to establish an oracle regarding the impact of performance regressions on the system-level performance, which is crucial for evaluating the effectiveness of our approach.
We examine the design documentation of our subject systems and identify typically triggered usage scenarios to exercise various aspects of the system. We choose the workload intensity that exercises the system performance without saturation since our approach is intended to work before extreme performance degradation.
In particular, for the TeaStore application, we have designed a diverse workload scenario in which the clients check the store, log into the system, and browse various categories and products, with the intensity of 50 requests/sec, and for the TrainTicket subject, we have realized a workload scenario in which the admin logs into the system, updates user information, and clients check their order and change a ticket reservation, with the intensity of 133 requests/sec. 
To ensure steady-state performance, we conduct 15-minute system performance testing for each run, in which we exclude the initial 5-minute warm-up period data and retain the subsequent 10-minute data as steady-state performance data. 
Furthermore, we repeat each system performance testing run three times to ensure robust results and mitigate the impact of the environmental noise and randomness during performance measurements.

\subsection{Architectural models}
\label{architectural_models}
Our study leverages architectural models of the system to detect system performance regressions, however, after manually checking both of our subject systems, we find that their architectural performance models are not explicitly provided.
Therefore, we need to construct the architectural models of our subject systems for evaluation purposes.
To achieve it, we rely on the collected system runtime behavior and performance information during system performance testing (cf. Section~\ref{test_design}) to construct the system's structure and calculate the model service demands, which is similar to the process in prior studies~\cite{DBLP:journals/entcs/ShoaibD11, DBLP:journals/tse/Kounev06}.
In our study, we utilize the QPN~\cite{DBLP:conf/pnpm/Bause93} modeling formalism to construct the architectural model due to its superior expressiveness and effectiveness in modeling the performance of complex software systems~\cite{DBLP:conf/cmg/TiwariM06}.
The architectural models used in our study are also shared in the replication package.
It is also worth noting that our study does not rely on specific types of architectural models, and the construction of architectural models is also not the primary focus of our study. Practitioners are encouraged to refer to the existing literature (cf. Section~\ref{sec:background}) for insights into various modeling techniques and methodologies.

\subsection{Evaluation scenario design}
\label{evaluation_deisgn}
We consider two evaluation scenarios to assess the effectiveness of our approach.
\revised{RCC7}{In the traditional in-house performance testing scenario, developers often fix the workloads when they model, test, and analyze the performance of different versions of systems~\cite{DBLP:conf/icst/GaoJBL16}.}
Therefore, we consider \textbf{fixed-workload} as our first evaluation scenario.
\revised{RAC4}{In this scenario, for each of our subject systems, we first perform system-level performance testing on both the original version (with no injected regression) and the updated versions (each version has an injected regression of a specific intensity and at a specific location in the source code) with our designed workload (cf. Section~\ref{test_design}).}
We then compare the performance testing results from the original and updated versions to understand the impact of performance regressions on the system performance in terms of response time and CPU percentage.
It is noteworthy that our approach does not inherently require system-level performance testing (cf. Section~\ref{sec:approach}), and the purpose of doing so is to establish an oracle regarding the impact of performance regressions on the system-level performance, which is crucial for evaluating the effectiveness of our approach.
Afterward, we apply our approach to each of the updated versions with the same workload and then examine whether our approach can detect performance regressions and keep consistent with the performance testing results.

In the second evaluation scenario, we assess whether our approach works when system workloads change between system versions. Changes in the system workload are common, as they are influenced by the dynamics of the operating environment (e.g., an increase in the number of users over time). 
In such cases, the same local performance deviation may impact the system-level performance differently. Therefore, we also consider a \textbf{various-workload} evaluation scenario to explore whether our approach can capture these diverse impacts and remain effective under various workloads.
\revised{RCC4}{In this scenario, we carefully devise three different workload variants based on the original workload. In particular, the first variant changes the original workload intensity, the second variant changes the original execution ratio among various request types, and the third variant combines both the first and second variants simultaneously.
}
Similar to the fixed-workload scenario, we first conduct system-level performance testing to understand the impact of the same regressions under various workloads on the system-level performance and then apply our approach to these various workloads.
We pay special attention to the minimum detectable regressions at each location in the fixed-workload scenario (if absent, then the highest intensity is used) to examine whether our approach can still successfully detect these performance regressions and remain consistent with the performance testing results under various workloads.

%% file: tex/Results.tex
\section{Evaluation Analysis and Results}
\label{sec:results}
\revised{RBC15}{In this section, we first present the data analysis approaches for the evaluation. Then, we discuss the evaluation results.}

\subsection{Data analysis}

Table~\ref{tab:fixed_workload_res} and Table~\ref{tab:various_workload_res} show the detailed results of applying our approach to detecting end-to-end system performance regressions for our subject systems in the \textbf{fixed-workload} and \textbf{various-workload} scenarios, respectively. We use the following three metrics to evaluate our approach.

\underline{MPD} refers to the mean performance degradation between two versions (with and without performance regressions). It explains how the regression impacts the mean of system performance. A positive value indicates a longer response time or higher CPU percentage in the updated version.

\underline{Effect size} is a statistical metric used to quantify the practical significance of the observed effects of performance regressions in a standardized way (independent of the measurement scales or units). 
It compares the performance distributions before and after injecting regressions.
This metric can help determine whether the observed effects are meaningful or merely coincidental, and its magnitude reflects effect strength.
We opt to calculate Cliff’s Delta effect size since it does not assume a specific distribution for the performance data.

\underline{Outcome} examines whether the impact of performance regressions measured during performance testing (i.e., the oracle) and the impact predicted by our approach are consistent. 
In particular, for response time, we consider the effect size larger than negligible as the presence of regressions, and the ``Outcome'' indicates the classification (including TP for true positive, TN for true negative, FP for false positive, and FN for false negative) of model prediction compared to performance testing based on the effect size, while for CPU percentage, since only mean CPU percentage is produced, the ``Outcome'' measures the absolute difference ($|\Delta|$) between the MPD of performance testing and model prediction.

\renewcommand{\arraystretch}{1.0}
\begin{table*}[!h]
\centering
\scriptsize
\tabcolsep=0.5em
\caption{Overall results of detecting end-to-end system performance regressions in the \textbf{fixed-workload} scenario}
\vspace{-1em}
\label{tab:fixed_workload_res}
\begin{subtable}[h]{1\textwidth}
\centering
    \begin{tabular}{ccrrrrcrrr}
    \toprule
    \multicolumn{2}{c}{\textbf{Local performance deviation}} & \multicolumn{7}{c}{\textbf{System performance regression}} \\

    \cmidrule(lr){1-2} \cmidrule(lr){3-10}
    
   \multirow{3}[1]{*}{\textbf{Location}} & \multirow{3}[1]{*}{\textbf{Intensity}} & \multicolumn{5}{c}{\textbf{Response time}} & \multicolumn{3}{c}{{\textbf{CPU percentage}}} \\
   
    & & \multicolumn{2}{c}{{\textbf{Performance testing}}} & \multicolumn{2}{c}{\textbf{Model prediction}} & \textbf{Outcome} & \multicolumn{1}{c}{\textbf{Performance testing}} & \multicolumn{1}{c}{\textbf{Model prediction}} & \textbf{Outcome ($|\Delta|$)}\\
   
    & & \textbf{MPD (ms)} & \textbf{Effect size} & \textbf{MPD (ms)} & \textbf{Effect size} & & \multicolumn{1}{r}{\textbf{MPD}}  & \multicolumn{1}{r}{\textbf{MPD}} &  \\
    
    \cmidrule(lr){1-2} \cmidrule(lr){3-7} \cmidrule(lr){8-10}
    
    \multirow{3}[1]{*}{L1} & Low & 7.02 & medium & 7.60 & small & TP & 6.53 & 6.09 & 0.44\\
    & Medium & 41.68 & large & 37.67 & large & TP & 33.03 & 28.82 & 4.21\\
    & High & 250.37 & large & 749.01 & large & TP & 110.60 & 149.55 & 38.95\\
    
    \cmidrule(lr){1-2} \cmidrule(lr){3-7} \cmidrule(lr){8-10}
    
    \multirow{3}[1]{*}{L2} & Low & 5.31 & p $>$ 0.05 & 0.39 & negligible & TN & 0.13 & 0.27 & 0.14\\
     & Medium & 4.52 & negligible & 1.36 & negligible & TN & 1.01 & -0.08 & 1.09\\
     & High & 8.65 & small & 6.42 & small & TP & 3.91 & 5.21 & 1.30\\
     
    \cmidrule(lr){1-2} \cmidrule(lr){3-7} \cmidrule(lr){8-10}
    
    \multirow{3}[1]{*}{L3} & Low & -1.08 & negligible & 0.11 & negligible & TN & -0.12 & $< 0.01$ & 0.12\\
     & Medium & 0.52 & p $>$ 0.05 & 0.51 & negligible & TN & -0.19 & 0.31 & 0.50\\
     & High & -0.88 & negligible & 2.11 & negligible & TN & -0.11 & 1.68 & 1.79\\
    
    \bottomrule
    \end{tabular}
        \vspace{.5em}
\caption{TeaStore}
\vspace{-.5em}
\label{tab:fixed_workload_teastore}
\end{subtable}
\begin{subtable}[h]{1\textwidth}
\centering
    \begin{tabular}{ccrrrrcrrr}
    \toprule
    \multicolumn{2}{c}{\textbf{Local performance deviation}} & \multicolumn{7}{c}{\textbf{System performance regression}} \\

    \cmidrule(lr){1-2} \cmidrule(lr){3-10}
    
   \multirow{3}[1]{*}{\textbf{Location}} & \multirow{3}[1]{*}{\textbf{Intensity}} & \multicolumn{5}{c}{\textbf{Response time}} & \multicolumn{3}{c}{{\textbf{CPU percentage}}} \\
   
    & & \multicolumn{2}{c}{{\textbf{Performance testing}}} & \multicolumn{2}{c}{\textbf{Model prediction}} & \textbf{Outcome} & \multicolumn{1}{c}{\textbf{Performance testing}} & \multicolumn{1}{c}{\textbf{Model prediction}} & \textbf{Outcome ($|\Delta|$)} \\
   
    & & \textbf{MPD (ms)} & \textbf{Effect size} & \textbf{MPD (ms)} & \textbf{Effect size} & & \multicolumn{1}{r}{\textbf{MPD}}  & \multicolumn{1}{r}{\textbf{MPD}} &  \\
    
    \cmidrule(lr){1-2} \cmidrule(lr){3-7} \cmidrule(lr){8-10}
    
    \multirow{3}[1]{*}{L1} & Low & 3.63 & p $>$ 0.05 & -0.25 & p $>$ 0.05 & TN &  0.41 & -0.03 & 0.44\\
     & Medium & -1.61 & p $>$ 0.05 & 1.01 & negligible & TN &  $<0.01$ & 0.28 & 0.28\\
     & High & 12.51 & small & 6.37 & negligible & FN & 1.97 & 1.69 & 0.28\\
     
    \cmidrule(lr){1-2} \cmidrule(lr){3-10}

    \multirow{3}[1]{*}{L2} & Low & 0.48 & small & 0.33 & negligible & FN & 1.16 & 0.89 & 0.27\\
     & Medium & 1.47 & medium & 1.64 & small & TP & 3.77 & 4.25 & 0.48\\
     & High & 5.34 & large & 8.00 & large & TP & 13.81 & 20.12 & 6.31\\

    \cmidrule(lr){1-2} \cmidrule(lr){3-10}

    \multirow{3}[1]{*}{L3} & Low & 0.04 & negligible & 0.46 & negligible & TN & 0.08 & 1.18 & 1.10\\
     & Medium & 0.40 & negligible & 1.72 & negligible & TN & 1.05 & 4.43 & 3.38\\
     & High & 1.86 & large & 9.12 & large & TP & 4.92 & 22.76 & 17.84\\
    \bottomrule
    \end{tabular}%
        \vspace{.5em}
\caption{TrainTicket}
\label{tab:fixed_workload_trainticket}
\end{subtable}
\vspace{-1em}
\begin{tablenotes}
    \item Note 1: ``MPD'' refers to the mean performance degradation between two versions (with and without performance regressions).
    \item Note 2: For response time, the ``Outcome'' indicates the classification of model prediction compared to performance testing based on effect size, while for CPU percentage, since only mean CPU percentage is produced, the ``Outcome'' measures the absolute difference ($|\Delta|$) between the MPD of performance testing and model prediction.
    \item Note 3: ``TP'' (true positive) and ``TN'' (true negative) indicate the successfully detected presence and absence of performance regressions by our approach, respectively, while ``FP'' (false positive) and ``FN'' (false negative) indicate falsely detected presence and absence of performance regressions, respectively.
\end{tablenotes}
\vspace{-1.7em}
\end{table*}

\renewcommand{\arraystretch}{1}
\begin{table*}[!h]
\centering
\scriptsize
\tabcolsep=0.5em
\caption{Overall results of detecting end-to-end system performance regressions in the \textbf{various-workload} scenario}
\vspace{-1em}
\label{tab:various_workload_res}
\begin{subtable}[h]{1\textwidth}
\centering
    \begin{tabular}{cccrrrrcrrr}
    \toprule
    \multicolumn{2}{c}{\textbf{Local performance deviation}} &  & \multicolumn{8}{c}{\textbf{System performance regression}} \\

    \cmidrule(lr){1-2} \cmidrule(lr){3-11}

    \multirow{3}[1]{*}{\textbf{Location}} & \multirow{3}[1]{*}{\textbf{Intensity}} & \multirow{3}[1]{*}{\textbf{Workload}} & \multicolumn{5}{c}{\textbf{Response time}} & \multicolumn{3}{c} {\textbf{CPU percentage}} \\
   
    & & & \multicolumn{2}{c}{\textbf{Performance testing}} & \multicolumn{2}{c}{\textbf{Model prediction}} & \textbf{Outcome} & \multicolumn{1}{c}{\textbf{Performance testing}} & \multicolumn{1}{c}{\textbf{Model prediction}} & \textbf{Outcome ($|\Delta|$)}\\
    
    & & & \textbf{MPD (ms)} & \textbf{Effect size} & \textbf{MPD (ms)} & \textbf{Effect size}  & & \textbf{MPD} &  \multicolumn{1}{r}{\textbf{MPD}} &  \\

    \cmidrule(lr){1-2} \cmidrule(lr){3-3} \cmidrule(lr){4-8} \cmidrule(lr){9-11} 
    
    \multirow{4}[1]{*}{L1}  & \multirow{4}[1]{*}{Low} & Original & 7.02 & medium & 7.60 & small & TP & 6.53 & 6.09 & 0.44 \\   
     &  & Variant 1 & 7.25 & small & 8.51 & small & TP & 12.24 & 12.29 & 0.05 \\
     &  & Variant 2 & 7.27 & medium & 7.58 & small & TP & 4.17 & 4.15 & 0.02 \\
     &  & Variant 3 & 7.22 & small & 7.90 & negligible & FN & 8.48 & 8.22 & 0.26 \\
     
    \cmidrule(lr){1-2} \cmidrule(lr){3-3} \cmidrule(lr){4-8} \cmidrule(lr){9-11} 
    
     \multirow{4}[1]{*}{L2}  & \multirow{4}[1]{*}{High} & Original & 8.65 & small & 6.42 & small & TP & 3.91 & 5.21& 1.30 \\
     &  & Variant 1 & 2.03 & small & 6.65 & small & TP & 6.24 & 10.59 & 4.35 \\
     &  & Variant 2 & 4.14 & small & 6.46 & small & TP & 4.28 & 7.10 & 2.82 \\
     &  & Variant 3 & -1.82 & negligible & 7.04 & small& FP  & 7.99 & 14.09 & 6.10 \\

    \cmidrule(lr){1-2} \cmidrule(lr){3-3} \cmidrule(lr){4-8} \cmidrule(lr){9-11} 
    
    \multirow{4}[1]{*}{L3}  & \multirow{4}[1]{*}{High} & Original & -0.88 & negligible & 2.11 & negligible & TN & -0.11 & 1.68 & 1.79 \\
     &  & Variant 1 & -3.16 & negligible & 2.08 & negligible & TN & 0.66 & 3.52 & 2.86 \\
     &  & Variant 2 & -2.73 & negligible & 2.06 & negligible & TN & 0.31 & 3.54 & 3.23 \\
     &  & Variant 3 & -1.81 & negligible & 2.16 & negligible & TN & 1.61 & 7.06 & 5.45 \\
    \bottomrule
    \end{tabular}
    \vspace{.5em}
\caption{TeaStore}
\label{tab:various_workload_teastore}
\vspace{-.5em}

\end{subtable}
\begin{subtable}[h]{1\textwidth}
\centering
\begin{tabular}{cccrrrrcrrr}
    \toprule
    \multicolumn{2}{c}{\textbf{Local performance deviation}} &  & \multicolumn{8}{c}{\textbf{System performance regression}} \\

    \cmidrule(lr){1-2} \cmidrule(lr){3-11}

    \multirow{3}[1]{*}{\textbf{Location}} & \multirow{3}[1]{*}{\textbf{Intensity}} & \multirow{3}[1]{*}{\textbf{Workload}} & \multicolumn{5}{c}{\textbf{Response time}} & \multicolumn{3}{c} {\textbf{CPU percentage}} \\
   
    & & & \multicolumn{2}{c}{\textbf{Performance testing}} & \multicolumn{2}{c}{\textbf{Model prediction}} & \textbf{Outcome} & \multicolumn{1}{c}{\textbf{Performance testing}} & \multicolumn{1}{c}{\textbf{Model prediction}} & \textbf{Outcome ($|\Delta|$)}\\
    
    & & & \textbf{MPD (ms)} & \textbf{Effect size} & \textbf{MPD (ms)} & \textbf{Effect size}  & & \textbf{MPD} &  \multicolumn{1}{r}{\textbf{MPD}} &  \\

    \cmidrule(lr){1-2} \cmidrule(lr){3-3} \cmidrule(lr){4-8} \cmidrule(lr){9-11} 
    
    \multirow{4}[1]{*}{L1}  & \multirow{4}[1]{*}{High} & Original & 12.51 & small & 6.37 & negligible & FN & 1.97 & 1.69 & 0.28 \\    
     &  & Variant 1 & 32.79 & small & 7.47 & negligible & FN & 5.31 & 2.80 & 2.51 \\
     &  & Variant 2 & 3.73 & small & 6.80 & negligible & FN & 0.73 & 1.73 & 1.00 \\
     &  & Variant 3 & 11.64 & small & 7.50 & negligible & FN & 3.10 & 2.89 & 0.21 \\

    \cmidrule(lr){1-2} \cmidrule(lr){3-3} \cmidrule(lr){4-8} \cmidrule(lr){9-11} 
     
    \multirow{4}[1]{*}{L2}  & \multirow{4}[1]{*}{Medium} & Original & 1.47 & medium & 1.64 & small & TP & 3.77 & 4.25 & 0.48 \\
    &  & Variant 1 & 1.36 & medium & 1.67 & small & TP & 5.26 & 6.33 & 1.07 \\
    &  & Variant 2 & 1.31 & medium & 1.66 & small & TP & 5.05 & 6.03 & 0.98 \\
    &  & Variant 3 & 2.49 & medium & 1.71 & small & TP & 7.92 & 9.07 & 1.15 \\
    
    \cmidrule(lr){1-2} \cmidrule(lr){3-3} \cmidrule(lr){4-8} \cmidrule(lr){9-11}      
    
    \multirow{4}[1]{*}{L3}  & \multirow{4}[1]{*}{High} & Original & 1.86 & large & 9.12 & large & TP & 4.92 & 22.76 & 17.84 \\
     &  & Variant 1 & 1.77 & medium & 9.71 & large & TP & 6.76 & 34.17 & 27.41 \\
     &  & Variant 2 & 1.74 & medium & 8.91 & large & TP & 3.20 & 16.17 & 12.97 \\
     &  & Variant 3 & 1.90 & medium & 9.18 & large & TP & 5.29 & 24.22 & 18.93 \\
    \bottomrule
    \end{tabular}
        \vspace{.5em}
\caption{TrainTicket}
\label{tab:various_workload_trainticket}
\end{subtable}
\vspace{-1em}
\begin{tablenotes}
    \item Note 1: ``MPD'' refers to the mean performance degradation between two versions (with and without performance regressions).
    \item Note 2: For response time, the ``Outcome'' indicates the classification of model prediction compared to performance testing based on effect size, while for CPU percentage, since only mean CPU percentage is produced, the ``Outcome'' measures the absolute difference ($|\Delta|$) between the MPD of performance testing and model prediction.
    \item Note 3: ``TP'' (true positive) and ``TN'' (true negative) indicate the successfully detected presence and absence of performance regressions by our approach, respectively, while ``FP'' (false positive) and ``FN'' (false negative) indicate falsely detected presence and absence of performance regressions, respectively.
\end{tablenotes}
\vspace{-1.7em}
\end{table*}

\subsection{Evaluation results}

\noindent\textbf{Evaluation in the fixed-workload scenario}:

\textbf{When there exist significant local performance deviations, their impact on end-to-end system performance may not always be significant.}
For example, as shown in Table~\ref{tab:fixed_workload_res}, in the TeaStore subject, for the local performance deviation at location L2 with Low and Medium intensities and L3 with all three intensities (i.e., Low, Medium, and High), there exist significant local performance deviations with up to 250\% slowdown (cf. Section~\ref{performance_regressions}), however, their impact on end-to-end system performance is all insignificant, i.e., with a p-value $> 0.05$ or only negligible effect size in response time and very low MPD ($\leq$ 1.01) in CPU percentage.
After further investigation, we observe that these components exhibiting performance deviations have an inherently low impact in terms of system performance. For instance, even when subjected to the most severe performance regression (i.e., injecting an extra 250\% waiting time), the execution time of these components mostly does not exceed 1 millisecond. Furthermore, these components are often executed infrequently during system runtime (e.g., only once), further diminishing their impact on system performance.
Such a finding indicates that significant local performance deviations may not always cause a notable impact at the system level, potentially leading to false positives. Therefore, if developers consistently choose to conduct system performance testing whenever there are notable local performance deviations, it could result in a considerable waste of resources and time. This finding signifies the importance of our approach.

\textbf{Our approach does not generate false alarms when there is local performance deviation but no system performance regression.}
As shown in Table~\ref{tab:fixed_workload_res}, when there is local performance deviation but no performance regression in system performance testing, our approach does not generate false alarms (i.e., FP). For instance, in TeaStore, for the local performance deviation at the location of L2 with Low and Medium intensities and L3 with all three intensities (i.e., Low, Medium, and High), when there exists insignificant system performance impact observed from performance testing, the results predicted from our approach always indicate a negligible difference between the system response times before and after injecting the performance regressions. 
In addition, regarding CPU percentage, for both of our subject systems (i.e., TeaStore and TrainTicket), the MPD between our model predictions and performance testing results are also similar. The largest difference between the model predicted and the actual MPD for CPU percentage is only 3.38 when there exist no system performance regressions.

\textbf{Our approach can effectively detect end-to-end system performance regressions caused by local performance deviations in the components.}
As presented in Table~\ref{tab:fixed_workload_res}, for the cases where the local performance deviation leads to notable system performance regression (i.e., with a small to large effect size in performance testing), our approach can detect seven out of nine cases in our subject systems with comparable results (in MPD and effect size) to the performance testing results in terms of response time.
The two (FN) cases missed by our approach are all from TrainTicket, i.e., L1-High and L2-Low. It is worth noting that despite missing these two cases, their actual impact on the system performance is relatively low (i.e., both have small effect sizes in performance testing).
\revised{RAC8, RBC9}{Furthermore, we suspect that one potential reason for these cases could be the quality of the architectural models that we recovered through the system-level performance testing (cf. Section~\ref{architectural_models}),
thus impacting the effectiveness of detecting performance regressions.}
Concerning CPU percentage, we observe that the differences between the model predicted and the actual MPD when there are system performance regressions are relatively larger than those when there are no regressions. However, such differences are consistently in the right direction, i.e., model prediction shows a larger MPD than performance testing, therefore, they do not fundamentally impact the determination of performance regressions.

\findingboxx{The presence of local performance deviations may not always indicate a notable deviation of the end-to-end system performance, thus conducting system performance testing in such cases would lead to wastage in terms of both time and resources. Our approach can successfully detect end-to-end system performance regressions without the execution of time- and resource-consuming system-level performance testing.}

\noindent\textbf{Evaluation in the various-workload scenario:}

\textbf{Our approach can maintain its effectiveness in detecting end-to-end system performance regressions even when the system experiences various workloads.}
\revised{RBC11}{As shown in Table~\ref{tab:various_workload_res}, we find that our approach can demonstrate effectiveness when the system experiences various workloads.}
In particular, for TrainTicket, except for the local performance deviation of L1, which is undetectable even under the original workload, the model prediction results consistently align with the performance testing results. As for TeaStore, our approach maintains consistency with the performance testing results in seven out of nine cases.
We notice an FN case in TeaStore for local performance deviation of L1-Low under variant 3, i.e., small regression in performance testing while negligible regression in model predictions. Upon a closer examination, we find that the MPD in response time from performance testing and model prediction is quite close, with values of 7.22 ms compared to 7.90 ms. In addition, the effect size for the model prediction results is 0.141, which is just slightly different from the threshold of small effect size at 0.147.
In addition, we identify an FP case in TeaStore at L2-High under variant 3, i.e., negligible regression in performance testing but small regression in model predictions. After further investigation, we find that from the performance testing results, the mean response time after injecting the performance regression is even slightly lower (i.e., with a negative MPD), which indicates a faster execution compared to the pre-injection state.
Such a phenomenon would be explained by the nature of the system's intricacies, where the system may cache certain hotspot codes or adopt other optimization strategies to improve overall performance. Nevertheless, capturing such optimizations in models can prove challenging and requires more consideration of the underlying structure and mechanisms of the system, which we leave for further research.

\textbf{Component-level performance testing can hardly capture the variety in the system workloads, thus it cannot consistently reflect the true effect of local performance deviation on the end-to-end system performance under various workloads.}
From Table~\ref{tab:various_workload_res}, we also observe that even the local performance deviation with the same intensity and at the same location in the source code can lead to different impacts (different MPD and effect size) on the end-to-end system performance when the system is under various workloads.
For instance, in the performance testing results of TeaStore, for the local performance deviation L1-Low, when the system experiences the workload variant 2, the MPD of CPU percentage is only 4.17. However, when the system experiences workload variant 1 (cf. Section~\ref{evaluation_deisgn}), the impact of local performance deviation on CPU percentage drastically escalates to 12.24 in MPD, approximately tripling the original value.
Upon deeper investigation, it has come to our attention that due to the increased workload intensity in variant 1, the component experiencing performance deviations is being executed much more frequently. As a result, the workload variant (i.e., variant 1) has resulted in a more significant performance impact at the system level, even with the same local performance deviation in the source code.
Such a finding implies that only focusing on the local performance deviation without considering the various system workloads is often not enough to accurately reflect the true effect of local performance deviation on the end-to-end system performance.

\findingboxx{Without considering the system workloads, it is often challenging to reflect the true effect of local performance deviations on the system performance. Our approach remains effective in detecting end-to-end system performance regressions even when the systems experience various workloads.}

%% file: tex/RelatedWork.tex
\section{Related Work}
\label{sec:relatedwork}

In this section, we discuss the prior research that is related to our work in two aspects, including:
1) detecting system performance regressions,
and 2) reducing the resources needed for performance testing.

\noindent \textbf{Detecting system performance regressions.}
Extensive prior research has proposed to utilize statistical analysis techniques~\cite{DBLP:conf/icsm/ChenS17, DBLP:conf/icse/MalikHH13, DBLP:conf/apsec/NguyenAJHNF11, DBLP:conf/icse/DingCS20, DBLP:conf/wosp/HegerHF13} to analyze the system performance testing results. 
For example, Nguyen et al.~\cite{DBLP:conf/apsec/NguyenAJHNF11} propose to utilize statistical process control charts to compare the performance metrics generated by the performance testing of both old and new version systems.
Since performance testing data is typically large in quantity and highly complex, existing research~\cite{DBLP:conf/icse/FooJAHZF15, DBLP:conf/seke/CellierDFR11, DBLP:conf/sigsoft/HeLLZLZ18, DBLP:conf/icsm/SyerJNHNF13} has also proposed to leverage data mining techniques to capture the existence of performance regressions. 
For instance, Foo et al.~\cite{DBLP:conf/icse/FooJAHZF15} propose to extract the association rules between multiple performance metrics. The changes in association rules may indicate performance regressions in the systems.
Additionally, machine learning models are also widely adopted in prior work~\cite{DBLP:conf/wosp/ShangHNF15, DBLP:conf/wosp/XiongPZG13, DBLP:conf/wosp/DidonaQRT15, DBLP:conf/icdm/LimLZFTLDZ14, DBLP:conf/dsn/ZhangCGSF05} to identify changes in system behavior that result in performance regressions.
For example, Zhang et al.~\cite{DBLP:conf/dsn/ZhangCGSF05} propose to build multiple machine learning models and utilize the ensemble algorithm to fuse the information from these models to adapt to the impact of changing workloads or external disturbances.

However, these approaches mainly detect system performance regressions based on time- and resource-consuming system performance testing. In contrast, our approach focuses on providing developers with an early impression of performance regressions, instead of running expensive system performance testing.

\noindent \textbf{Reducing the resources needed for performance testing.}
\revised{RBC16}{Along with the component-level performance testing discussed in Section~\ref{sec:background}, extensive research has also explored various other techniques to reduce the resources (e.g., time) needed for performance testing.}
In particular, prior studies~\cite{DBLP:conf/icsm/AlghmadiSSH16, DBLP:conf/icse/BusanyM16, DBLP:conf/sigsoft/HeMS0PS19} propose to utilize various statistical analysis techniques to measure whether the performance metrics or log traces collected during performance testing are repetitive and searching for the point to stop performance testing early.
In addition, there are also prior studies~\cite{DBLP:conf/icse/Shariff0BHNF19, DBLP:journals/tse/GranoLPP21} proposing to reduce the system resources during testing a workload. For example, Shariff et al.~\cite{DBLP:conf/icse/Shariff0BHNF19} focus on browser-based (with Selenium) performance testing. They propose to share the browser instances between the simulated test user instances, thereby reducing the total number of browser instances and the system resources needed during testing.

Prior research mainly reduces the time and resources of system performance testing or proposes to utilize the tests on a smaller scale. However, system performance testing is conducted late, making the diagnosis challenging and laborious, while component-level performance testing does not explain the system-level performance well. 
In comparison, in our work, we bridge local performance data and architectural models earlier without the need to run expensive system performance testing after the system is built or even integrated.

%% file: tex/Threats.tex
\section{Threats to Validity}
\label{sec:threats}

This section discusses the threats to the validity of our study.

\revised{RBC4, RCC2}{
\noindent \textbf{External validity.}
Our study is performed on two representative and prevalent open-source systems.
Although both subject systems adopt the microservice architecture, it is worth noting that our approach is not limited to the unique characteristics of microservices, but rather applicable to all systems with subsystems. 
Nevertheless, more case studies on other software systems adopting other types of architecture and in other domains can benefit the evaluation of our approach.
In addition, our study only considers one type of architectural model (i.e., QPN model) developed from performance testing data (cf. Section~\ref{architectural_models}). However, in practice, there exist other types of architectural performance models (e.g., the LQN model), and the models can also be developed from other sources, like actual field performance data. Considering more types of architectural models developed from various sources of performance data is in our future plan.
}

\revised{RBC4}{
\noindent \textbf{Internal validity.}
Our study relies on the performance data collected during the component-level and system-level performance testing of the system. The unsatisfactory quality of the performance data, e.g., non-stable or high-noisy data, can adversely impact the internal validity of our study.
To mitigate this threat, we conduct our experiments with rigorous measures. We repeat the experiment 30 times for local performance testing, and for system performance testing, we repeat the experiment 3 times, each lasting 15 minutes. Additionally, we exclude data from the initial 5-minute warm-up period. %
Our approach also depends on the availability and the accuracy of an architectural model of the software system. Therefore, our approach may not be feasible when there is no architectural model and may not perform well when the model provides poor accuracy or the model is outdated.
}

\revised{RBC5}{
\noindent \textbf{Construct validity.}
Our study is conducted under constant load during system performance testing, and the system behaviors may differ when conducted under continuously varying loads. %
Future work may extend our approach under continuously varying loads. 
}

\revised{RBC3, RBC5}{
\noindent \textbf{Conclusion validity.}
Our evaluation results depend on thresholds to determine the statistical significance, while in practice, developers usually rely on their subjective judgment to determine the significance for their systems.
}

%% file: tex/Conclusion.tex
\section{Conclusion}
\label{sec:conclusion}

In this paper, we propose a novel approach to detecting end-to-end system performance regressions as early as in the software development phase, i.e., before the expensive system performance testing.
By bridging the local performance data collected during the performance testing of individual components and the architectural models of the entire system, our approach can predict the impact of component-level performance deviations on the end-to-end system performance, enabling lightweight performance regression detection.
Our evaluation results on two representative benchmark systems demonstrate that our approach can effectively detect system performance regressions with different intensities of local performance deviations and under various system workloads.
Furthermore, our proposed approach also offers a promising solution to complement traditional system performance testing to ensure the software performance in the fast-paced software development and release practice (e.g., DevOps) where testing resources are limited.